\documentstyle[twocolumn,prb,aps,epsf]{revtex}
\begin{document}
\draft

\title
{Relation between Vortex core charge
and Vortex Bound States}

\author{
Nobuhiko Hayashi, Masanori Ichioka, and Kazushige Machida
}
\address{
Department of Physics, Okayama University, Okayama 700-8530,
Japan}
\date{Submitted July 15, 1998; to be published
in J.~Phys.~Soc.~Jpn.~{\bf 67}, No.~10 (1998)}
\maketitle

\begin{abstract}
   Spatially inhomogeneous electron distribution around a single vortex
is discussed on the basis of the Bogoliubov-de Gennes theory.
   The spatial structure and temperature dependence
of the electron density around the vortex are presented.
   A relation between the vortex core charge and the vortex bound states
(or the Caroli-de Gennes-Matricon states)
is pointed out.
   Using the scanning tunneling microscope,
information on the vortex core charge can be extracted
through this relation.
\end{abstract}


   Electric charging phenomena around vortices have the potential of
becoming one of the key features in the physics of
the mixed state in type-II superconductors.
   Until quite recently, little was known about the electric charging
inherent in vortices,
while it has been well recognized since the 1950's that
each vortex line carries
a quantized magnetic flux.
   Only recently, it was noticed that an electric charge
accumulates around a static vortex line in type-II superconductors.
   Khomskii and Freimuth,~\cite{khomskii}
and Blatter {\it et al.}~\cite{blatter}
theoretically discussed
the electric charging around a vortex.~\cite{topol}
   If the electric charging of vortices is experimentally confirmed,
it will open the door to an unexplored field in which one expects various
electromagnetic phenomena to originate from the electric charge trapped by
vortices.

   In spite of the growing interest in vortex core charging,
firm experimental evidence of the charging is lacking
at the present.
   However, various experimental attempts are now in progress
and are on the verge of detecting a charge accumulation inside vortex cores.
   One such experiment is a polarized neutron scattering investigation
of the flux line lattice in Nb by Neumann {\it et al.}~\cite{neumann}
   They detected a nonzero nuclear contribution to the Bragg peaks
corresponding to the periodicity of the flux line lattice.
   This experimental result strongly suggests existence of
the vortex core charge;
if each vortex constituting the flux line lattice traps electrons or holes,
the response of the underlying nuclei to these will induce a distortion in
the nuclear lattice around the vortices.~\cite{neumann,kusmartsev}
   Various types of experiments which will attempt to
detect the vortex core charge
are also planned.
   In addition, an experiment to observe the temperature $T$
and magnetic field $H$ dependence of the vortex core charge
is expected
in order to
establish the existence of the vortex core charging.
   Therefore, it is certainly desired that detailed theoretical predictions
for the temperature or magnetic field dependence of the vortex core charge
should be presented for experimental verification.

   In this paper, we present the structure of the carrier density
around a static single vortex
and its temperature dependence, solving self-consistently
the Bogoliubov-de Gennes (BdG) equation.
   On the basis of the solution of the BdG equation,
we discuss not only the temperature dependence but also
the relation between the charging of the vortex core and
the so-called Caroli-de Gennes-Matricon (CdGM) states
(or the vortex bound states).

   The CdGM states, i.e.,
low-energy excited states due to vortices,~\cite{haya-prb}
were first discussed theoretically by Caroli {\it et al.}~\cite{caroli}
   Their existence was experimentally confirmed
by Hess {\it et al.},~\cite{hess}
who observed spatial dependence of the excitation spectra
around a vortex with scanning tunneling microscopy (STM).
   The local density of states (LDOS) around a vortex,
probed by STM, depends on
the Bogoliubov wave functions
of the CdGM states $u_j({\bf r})$ and $v_j({\bf r})$,
labeled by the quantum number $j$.
   The LDOS $N({\bf r},E)$ (to be exact, thermally smeared LDOS, i.e.,
the tunneling conductance) is given as
\begin{equation}
N({\bf r},E) = \sum_j \bigl[|u_j({\bf r})|^2 f'(E-E_j)
+|v_j({\bf r})|^2 f'(E+E_j) \bigr],
\label{eq:1}
\end{equation}
where $E_j$ is the eigenenergy and $f(E)$ the Fermi function
(the prime represents the derivative).
   The STM enables us to extract detailed information on
the wave functions around a vortex.
   Here, we notice that the carrier density around a vortex,
$n({\bf r})$, also depends on these wave functions:
\begin{equation}
n({\bf r}) = 2\sum_{E_j >0} \bigl[|u_j({\bf r})|^2 f(E_j)
+|v_j({\bf r})|^2 \{1-f(E_j)\} \bigr].
\label{eq:2}
\end{equation}
   The electric charging (or the inhomogeneous electron density distribution)
around a vortex is related to the LDOS through the wave functions
$u_j({\bf r})$ and $v_j({\bf r})$. 
   This suggests unique potential ability
of the STM; the structure of the LDOS probed by STM relates to
the spatial structure of the vortex core charge.

   Regarding the previous theories of the mechanism of
the vortex core charging,
Khomskii and Freimuth~\cite{khomskii}
based their scenario on a normal-core model.
   Assuming that the vortex core is a region of normal metal surrounded
by a superconducting material,
they considered that the corresponding difference
in the chemical potential~\cite{khomskii2}
leads to a redistribution of the electrons.~\cite{khomskii}
   Blatter {\it et al.}~\cite{blatter}
discussed the charging mechanism,
considering spatial variation of the pair potential $\Delta({\bf r})$
around a vortex.
   On the basis of the zero-temperature version of eq.~(\ref{eq:2}),
they obtained $n({\bf r})$ by
combining the spatial variation of the wave function
$v({\bf r})$ with particle-hole asymmetry in the normal-state
density of states at the Fermi level.
   The discussion was, however, based on a wave function which
had the same form as the {\it uniform solution} of the BdG equation,
namely~\cite{blatter}
\begin{eqnarray}
v_k({\bf r})=\sqrt{\frac{1}{2} \Bigl(1-\frac{\xi_k}{E_k}\Bigr)},\quad
E_k=\sqrt{\xi_k^2 + |\Delta({\bf r})|^2}.
\label{eq:3}
\end{eqnarray}
   The spatial variation of $v_k({\bf r})$ was directly determined
by the local value of $\Delta({\bf r})$,
which is not exactly appropriate for the vortex system.
   It is desired that one should base the calculation on the exact
wave functions
of the CdGM states.

   Prompted by this motivation,
we will self-consistently solve the BdG equation to obtain
the exact wave functions
$u_j({\bf r})$ and $v_j({\bf r})$ of the CdGM states
(including the extended states above the gap).
   We start with the BdG equation given, in a dimensionless form, by
\begin{eqnarray}
\biggl[\frac{-1}{2 k_{\rm F}\xi_0}\nabla^2-\mu \biggr] u_j({\bf r})+
\Delta({\bf r})v_j({\bf r})
=E_j u_j({\bf r}),
\nonumber
\end{eqnarray}
\begin{eqnarray}
-\biggl[\frac{-1}{2 k_{\rm F}\xi_0}\nabla^2-\mu \biggr] v_j({\bf r})+
\Delta^{\ast}({\bf r})u_j({\bf r})
=E_j v_j({\bf r}),
\label{eq:bdg}
\end{eqnarray}
where $\mu$ is the chemical potential and
$\xi_0$(=$v_{\rm F} / \Delta_0$) is the coherence length 
[$\Delta_0$ is the uniform gap at $T=0$,
and $k_{\rm F}$ ($v_{\rm F}$) is the Fermi
wave number (velocity)].
   In eq.~(\ref{eq:bdg}),
   the length (energy) scale is measured by $\xi_0$ ($\Delta_0$).
   For an isolated single vortex
in an extreme type-II superconductor,
we may neglect the vector potential in eq.~(\ref{eq:bdg}).
   To maintain macroscopic charge neutrality in the material,
in eq.~(\ref{eq:2}) we constrain the electron density in a uniform system
to be constant on the temperature.
   We use $\mu$ determined at each temperature by this constraint,
which is equivalent at zero temperature
to eq.~(4) of ref.~\onlinecite{khomskii2}.
   The pair potential is determined self-consistently by
\begin{eqnarray}
\Delta({\bf r})=
g\sum_{|E_j|\leq \omega_{\rm D}}
u_j({\bf r})v^{\ast}_j({\bf r})\{1-2f(E_j)\},
\label{eq:gap}
\end{eqnarray}
where $g$ is the coupling constant and $\omega_{\rm D}$ the energy cutoff,
which are related by the BCS relation
via the transition temperature $T_c$ and
the gap $\Delta_0$.
   We set $\omega_{\rm D}=20\Delta_0$.
   We consider, for clarity,
an isolated vortex under the following conditions.
(a) The system is a cylinder with a radius $R$.
(b) The Fermi surface is cylindrical.
(c) The pairing has isotropic $s$-wave symmetry.
   Thus the system has cylindrical symmetry.
   We write the eigenfunctions as
$u_j({\bf r})=u_{n,l}(r) \exp\bigr[i(l-\frac{1}{2})\theta \bigl]$ and
$v_j({\bf r})=v_{n,l}(r) \exp\bigr[i(l+\frac{1}{2})\theta \bigl]$ with
$\Delta({\bf r})=\Delta(r) \exp\bigr[-i\theta \bigl]$ 
in polar coordinates, where $n$ is the radial quantum number and 
the angular momentum $|l|=\frac{1}{2},\frac{3}{2},\frac{5}{2},\cdots$.
   We expand the eigenfunctions in terms of
the Bessel functions~\cite{caroli}
$J_m(r)$ as~\cite{gygi}
\begin{eqnarray}
u_{n,l}(r)=\sum_{i=1}^N c_{ni}\phi_{i|l-\frac{1}{2}|}(r),
\nonumber
\end{eqnarray}
\begin{eqnarray}
v_{n,l}(r)=\sum_{i=1}^N d_{ni}\phi_{i|l+\frac{1}{2}|}(r),
\label{eq:6}
\end{eqnarray}
where $\phi_{im}(r)=[{\sqrt 2}/ RJ_{m+1}(\alpha_{im})]
J_m(\alpha_{im}r/ R)$
and $\alpha_{im}$ is the $i$-th
zero of $J_m(r)$.
   We set $R=20\xi_0$.
   The BdG equation is reduced to a $2N\times 2N$
matrix eigenvalue problem.
   This useful technique to solve eq.~(\ref{eq:bdg}),
developed by Gygi and Schl\"uter,~\cite{gygi}
has been utilized in some cases.~\cite{gygi,tanaka,franz,iso,morita,haya}
   Our system is characterized by a parameter
$k_{\rm F}\xi_0$,~\cite{morita,haya}
important for the present problem.
   From our standpoint, all interactions between the quasiparticles are
renormalized to $g$ in eq.~(\ref{eq:gap})
and additional screening does not exist in the Hamiltonian.
   The screening for the charge ordering is excluded
as in the charge density wave studies.~\cite{cdw}
   If some screening effect is considered,
in principle we may take it into account as an external potential
in eq.~(\ref{eq:bdg}) and solve self-consistently the equations
together with
an additional equation, e.g., the Poisson's equation.
   Such a study, if meaningful, is left for a future work.
   Using the calculated $u_j({\bf r})$ and $v_j({\bf r})$,
we obtain the LDOS $N({\bf r},E)$ and the carrier density $n({\bf r})$
from eqs.\ (\ref{eq:1}) and (\ref{eq:2}), respectively.

\begin{figure}
\epsfxsize=65mm
\hspace{5.5mm}
\epsfbox{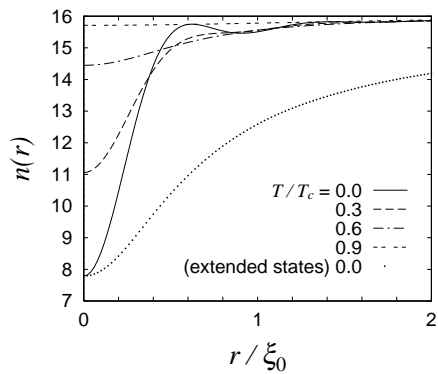}
\vspace{5.5mm}
\caption{
     The spatial variation of the carrier density $n(r)$
     (in arbitrary units) around the vortex for several temperatures
     and $k_{\rm F}\xi_0=4$.
     The contribution of the extended states to $n(r)$
     at $T=0$ is shown by the dotted line.
     The difference between this contribution and the total $n(r)$
     at $T=0$ gives the contribution of the bound states.
}
\label{fig:density}
\end{figure}

   In Fig.~\ref{fig:density}, we present
the spatial structure of the carrier density
$n(r)$ around the vortex at several temperatures.
   The Friedel oscillation appears at low temperatures, because
each wave function of the low-energy CdGM states oscillates with
a period $\simeq k_{\rm F}^{-1}$.
   It is striking that the carrier density at the vortex center exhibits
strong temperature dependence and leads to a substantial charging
at low temperatures.

   The carrier density at the vortex center
in Fig.~\ref{fig:density}
decreases
with respect to that far from the core.
Consequently,
in the case of the present electron system
(i.e., the two-dimensional free electron system),
the sign of the vortex core charge is opposite to
the sign of the electron
which is the dominant charge carrier in the present case.
   When the dominant charge carriers are holes,
we only have to treat these holes as carriers in that system
instead of the electrons
and there are no changes in the formulation
[eqs.~(\ref{eq:1}), (\ref{eq:2}), (\ref{eq:bdg})--(\ref{eq:6})].
   The density of the dominant carriers (holes) decreases
near the vortex center in this case as well.

\begin{figure}
\epsfxsize=65mm
\hspace{5.5mm}
\epsfbox{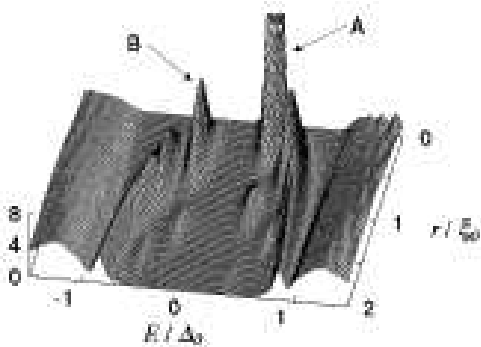}
\caption{
    The spectral evolution of the LDOS $N(E,r)$
    (in arbitrary units)
    at $T/T_c=0.05$ and $k_{\rm F}\xi_0=4$.
}
\label{fig:ldos}
\end{figure}

   The density of the dominant carriers decreases
near the vortex center,
as long as the wave functions
around a vortex for the dominant carriers
are given by eq.~(\ref{eq:6}).~\cite{oka}
   This is related to particle-hole asymmetry in the LDOS
{\it inside the vortex core}
and can be understood in connection with the CdGM states
as follows.
   In the definition of the angular momentum $l$ in eq.~(\ref{eq:6}),
the bound-state energy spectrum
is $E_l>0$ for $l>0$
(and $E_l<0$ for $l<0$), where $E_l=-E_{-l}$.
   In Fig.~\ref{fig:ldos}, we show the spectral evolution
obtained from eq.~(\ref{eq:1}).
   In systems where $k_{\rm F}\xi_0$ is small (the quantum limit),
the asymmetry in the LDOS
appears conspicuously.
   The two largest peaks near $E=0$ are noticeable
[the peaks $A$ and $B$].
   The peak $A$ at $E=E_{l=1/2}$ ($>0$)
is composed of $|u_{l=1/2}(r)|^2 \bigl(=|v_{l=-1/2}(r)|^2\bigr)$.
   The peak $B$ at $E=E_{-1/2}$ ($<0$)
is composed of $|v_{1/2}(r)|^2 \bigl(=|u_{-1/2}(r)|^2\bigr)$.
   From eq.~(\ref{eq:6}), $u_{1/2}(0)\neq 0$ and $v_{1/2}(0)=0$
because $J_m(0)\neq 0$ only for $m=0$.
   The asymmetry between $u_{1/2}(r)$ and $v_{1/2}(r)$
leads to the particle-hole asymmetry in the LDOS inside the core.~\cite{haya}
   Now, according to eq.~(\ref{eq:2}), $n(r)$ is constructed from
the wave functions which belong to $E>0$.
   The contribution from the extended states ($E>\Delta_0$)
is presented as the dotted line in Fig.~\ref{fig:density}.
   The remaining contribution to $n(r)$ comes from the bound states.
   The lowest bound state $v_{1/2}(r)$,
which belongs to the lowest bound state eigenenergy $E_{1/2}>0$,
predominantly determines the structure of $n(r)$
in the vicinity of the vortex center.
   The amplitude $|v_{1/2}(r)|^2$ is equal to that of the peak $B$
in the LDOS.
   The spatial profile of $n(r)$ is determined
by the shape of $|v_{1/2}(r)|^2$, i.e.,
the peak $B$.
   Since $|v_{1/2}(r)|^2$ decreases to zero with $r \rightarrow 0$
as seen from the spatial profile of the peak $B$ in Fig.~\ref{fig:ldos},
we can infer that $n(r)$ decreases near the vortex center.

   According to discussions~\cite{blatter} based on eq.~(\ref{eq:3}),
the carrier density near the vortex center
has a sensitive dependence on
the slope in the density of states.
   It might be expected that
if the derivative of the density of states
is negative,
the carrier density increases at the vortex center.
   To examine it,
we have investigated the case of the energy band,
$k^2/2m +k^4/4m^2 \varepsilon_0$,
(see ref.~\onlinecite{otterlo})
which has a negative derivative of the density of states
in two dimensions.
   In the calculation with a fixed $\mu$,
the density far from the core certainly
decreases with the growth of the gap $\Delta(T)$ on lowering $T$,
which is consistent with the precondition of ref.~\onlinecite{blatter}.
   In this situation,
on the basis of eq.~(\ref{eq:3}),
the density $n(r)$ is naively expected to recover
to the normal-state value on approaching the center $r=0$
where $\Delta(r)=0$.
$n(r)$ is then expected to increase at the center.
   However, according to results of the calculation
based on the wave functions of the CdGM states,
$n(r)$ decreases at the vortex center.
   We conclude that,
the carrier density near the vortex center
is determined by the electronic structure inside the vortex core,
which is insensitive to the slope
in the normal-state density of states at the Fermi level.

\begin{figure}
\epsfxsize=65mm
\hspace{5.5mm}
\epsfbox{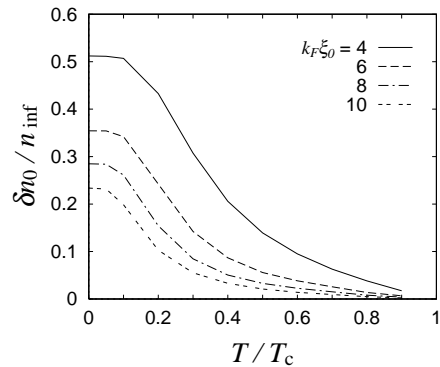}
\vspace{5.5mm}
\caption{
    The temperature dependence of the carrier density $n_0=n(r=0)$
    at the vortex center.
    In the figure
    $\delta n_0/n_{\infty}$ are plotted
    for several $k_{\rm F}\xi_0$,
    where
    $\delta n_0=|n_0-n_{\infty}|$ and
    $n_{\infty}$ ($\equiv n_{\rm inf}$)
is the plateau density far from the core.
}
\label{fig:temp-d}
\end{figure}

   Let us focus on the magnitude of the core charge.
   The carrier density at the vortex center,
from which the order of magnitude of the core charge is estimated,
exhibits substantial temperature dependence
as shown in Fig.~\ref{fig:temp-d}.
   We plot $\delta n_0/n_{\infty}$, where
$\delta n_0=|n_0-n_{\infty}|$, $n_0=n(r=0)$, and
$n_{\infty}$ is the plateau density $n_{\infty}=n(R/2)$,
to which the calculated $n(r)$
settles away from the core.
   We note that the $k_{\rm F}\xi_0$ dependence of the density,
$\delta n_0/n_{\infty}
\sim (k_{\rm F}\xi_0)^{-\alpha}
\simeq (\Delta_0/\varepsilon_{\rm F})^{\alpha}$,
varies with the temperature
($\varepsilon_{\rm F}$ is the Fermi energy).
   Our numerical data show that $\alpha \approx 1$ near $T=0$ and
$\alpha \approx 2$ near $T=0.5T_c$.
   The exponent $\alpha$ is crucial to
the magnitude of the core charge.
   In most conventional superconductors, 
the parameter $k_{\rm F}\xi_0$ is of the order of 100.
   It can be 1 -- 10 in high-$T_c$ cuprates.
   Depending on the estimate of $\alpha$,
   there can appear substantial difference in the evaluation of
the magnitude of the core charge.
   According to our results,
$\alpha$ depends on the temperature
as above.
   To estimate the total core charge $Q_v$ per unit length
along the vortex axis,
we consider the charging volume in Fig.~\ref{fig:density}
to be a cone with a height $\delta n_0$ and
a base radius $r_1$ ($2 k_{\rm F} r_1 = \pi$).
   $n(r)$ almost recovers to $n_{\infty}$ initially at
$r_1 \sim k_{\rm F}^{-1}$ at low temperatures.
   $Q_v$ is evaluated as $Q_v \approx e \pi r_1^2 \delta n_0 /3$.
   We consider a pancake vortex in a layer, and
the distance between each layer is $d$.
   In this case $n_{\infty} =2\pi k_{\rm F}^2 (2\pi/d)/8\pi^3$.
   We then obtain $Q_v \sim e (k_F\xi_0)^{-\alpha} d^{-1}$
at low temperatures.

   We should comment on the vortex dynamics
in the context of the above temperature dependence of $n_0$,
although the issue concerning the dynamics is
seriously controversial at the present time.~\cite{CandR}
   Feigel'man {\it et al.}~\cite{feigelman}
proposed a nondissipative transverse force acting on a vortex
originating from $\delta n_0$ (see also ref.~\onlinecite{otterlo}).
   Kopnin {\it et al.} reported that
the effect proposed by Feigel'man {\it et al.}~\cite{feigelman}
can be understood from the viewpoint of the spectral flow theory,
where $n_0$ is regarded
as the spectral flow parameter $C_0$.~\cite{spect,volovik}
   The parameter $C_0$ is independent of the temperature.
   Hence it appears to be inconsistent with the temperature dependence
of $n_0$ presented in this paper.
   Even in a neutral system with a fixed $\mu$,
$n_0$ exhibits substantial temperature dependence in our calculations.
   While Kopnin~\cite{kopnin} discussed the temperature dependence
of that force, the temperature dependence of $\delta n_0$ itself
at the vortex center seems not to be explicitly included there.
   We hope for a further investigation
based on the CdGM solutions~\cite{caroli}
to reveal possible mutual relations
between these theories
(refs.\ \onlinecite{feigelman,spect} and \onlinecite{kopnin})
and the significant temperature dependence of $n(r)$
in the present paper.~\cite{dziarmaga}

   We point out a relation between the present work and STM experiments.
   Maggio-Aprile {\it et al.}~\cite{maggio} and
Renner {\it et al.}~\cite{renner98}
observed spectral evolutions of the LDOS inside the vortex cores
in the high-$T_c$ cuprates.
   They detected particle-hole asymmetry in the LDOS near the core
center (see Fig.\ 2 in ref.~\onlinecite{renner98}).
   We expect that the asymmetry observed in the experiments
has the same origin as the asymmetry shown in our Fig.~\ref{fig:ldos}
(see also ref.~\onlinecite{haya}).
   We speculate that even if the superconductivity in the compounds
consists of the preformed pairs or 
is in the crossover region
between the BCS superconductivity and
the Bose-Einstein condensation,
the Bogoliubov wave functions would still be defined.
   If so, the electronic state of the vortex core
in the compounds is understood as the Andreev scattering~\cite{rainer}
and it is the coherent state.
   From our results based on the Bogoliubov wave functions,
we conclude that
the particle-hole asymmetry inside the vortex core
observed in the experiments~\cite{maggio,renner98}
implies the corresponding existence of
the vortex charging.
   According to another STM experiment
by Renner {\it et al.},~\cite{renner}
the coherent electronic structure inside the core,
observed as sharp structure of the LDOS,
is smeared gradually by impurity doping.
   We predict that the vortex core charge decreases by impurity doping,
because the charging is related to the sharp LDOS structure inside
the vortex core in our scenario.

   In summary, we investigated the electron density around a single
vortex on the basis of the BdG theory.
   Its temperature dependence was presented.
   We expect that experimental data regarded as the vortex core charge
will exhibit the temperature dependence as shown in Fig.~\ref{fig:temp-d}.
   If such dependence is observed, those experimental data
will become solid evidence of the vortex core charging.
   We discussed the microscopic charging mechanism,
{\it which is independent of the slope
in the normal-state density of states at the Fermi level},
by considering the CdGM states around the vortex.
   We pointed out the relation between the vortex bound states,
probed potentially by STM, and
the vortex core charging,
based on the inherent particle-hole asymmetry
inside the vortex core
{\it originating from the CdGM states of the vortex}.

   We are grateful to M.~Nishida, M.~Machida,
Y.~Matsuda, K.~R.~A.~Ziebeck,
T.~Isoshima and M.~Takigawa for useful discussions.
   N.H.\ and M.I.\ are supported by the Japan Society
for the Promotion of Science for Young Scientists.

\end{document}